\documentclass[preprint]{elsarticle}

\usepackage[bookmarks=false,hidelinks]{hyperref}

\graphicspath{{./imgs/}}

\usepackage{amssymb}
\usepackage{amsmath}

\usepackage{tabulary}

\usepackage[per-mode=symbol]{siunitx}
\DeclareSIUnit\g{g}

\usepackage{caption}

\biboptions{authoryear}

\usepackage[figuresright]{rotating}

\usepackage[acronym,nomain]{glossaries}
\makeglossaries
\newacronym{nmpc}{NMPC}{Non-linear Model Predictive Control}
\newacronym{ms}{MS}{Motion Sickness}
\newacronym{msi}{MSI}{Motion Sickness Incidence}
\newacronym{hmi}{HMI}{Human-Machine Interface}

\usepackage{pgfplots,tikz,tikzscale}
\pgfplotsset{compat=newest}

\begin{document}
\begin{frontmatter}

\title{Optimal speed profile on a given road for motion sickness reduction}

\author[UniFi]{Cesare Certosini}
\ead{cesare.certosini@unifi.it}
\author[UniFi]{Renzo Capitani}  
\author[M42]{Claudio Annicchiarico}

\cortext[cor1]{Corresponding author}

\address[UniFi]{Department of Industrial Engineering --- University of Florence, Florence, Italy}
\address[M42]{Meccanica 42 s.r.l., Sesto fiorentino (FI), Italy}

\begin{abstract}

\gls{ms} is an issue of most transportation systems.
Several countermeasures for such problem in cars are proposed in the literature, but most of them are qualitative, behavioural or involving complex chassis systems.
With the growing interest in self-employed vehicles, the issue of \gls{ms} may be so important that it undermines their benefits in terms of increased productivity; not addressing such issue may limit the users' acceptance reducing the safety and environmental impact of autonomous vehicles.
The present study presents a novel approach to combine minimal travel time with minimal \gls{msi} by optimising the speed profile for a given path.
Through simulation, a simple vehicle model is used to compare several strategies evaluating which of them are effective and which not.
The optimisation task is formulated as a \gls{nmpc} and a series of optimisation routines are computed along the path; the strategies are implemented within the cost function of the \gls{nmpc} problem evaluating their performance and if using a numerical \gls{ms} model is mandatory to get a significant reduction of the \gls{msi}.
The results show that not all the cost functions are effective, but it is possible to reduce \gls{ms} without modelling its dynamics; however, the strategy taking into account the current \gls{msi} using a \gls{ms} model outperforms the other cost functions in term of efficacy and efficiency.
Such a quantitative approach can be used during motion planning in autonomous vehicles, to suggest an optimal speed in human-driven vehicles and to improve the comfort for bus-line services or even trains.

\end{abstract}


\end{frontmatter}

\section{Introduction}

\gls{ms} is a known issue in transportation systems since ancient times.
The most famous versions of \gls{ms} are sea-sickness, car-sickness, air-sickness and space-sickness; however, in recent times simulator-sickness and virtual reality-sickness are different versions of the same issue with a growing interest for the scientific community.

The main theory explaining \gls{ms} is the conflict theory by \cite{reason_motion_1975}: the acceleration perceived conflicts with the expected one generating a conflict leading to \gls{ms}; this phenomenon is amplified when there is a conflict between the information coming from different sensorial paths: when a passenger in a car is reading it is more likely to be motion sick because of the visual information of the standing field of view conflicts with the vestibular system perceiving the actual acceleration of the vehicle, the reader should refer to \cite{bronstein_visual_2020} for a more detailed description of the visual-vestibular conflict.
\cite{young_visually_1978} shows that the visual-vestibular interaction is dominated by the visual information for frequencies below \SI{0.1}{\hertz}, while vestibular information dominates higher frequencies perception.
As \cite{ohanlon_motion_1973} shows, the most critical acceleration frequencies for \gls{ms} are the ones where there is no clear dominant path in the motion perception: the percentage of people that experienced emesis, called \gls{msi}, is maximum at \SI{\approx 0.17}{\hertz} and is significant for frequencies from \SIrange{0.01}{1}{\hertz}.

The research interest is very strong because such frequencies are characteristic for all transportation systems, so every kind of transportation system is affected by different versions of the same issue: sea-sickness was the first as per \cite{irwin_pathology_1881}, nevertheless, further studies have continued in more recent years and lead to the fundamental experimental studies of \cite{ohanlon_motion_1973,lawther_motion_1986}; airsickness is also studied as per \cite{turner_airsickness_2000}; space sickness is another variant of the same issue in a slightly different context and an example of the scientific interest is in \cite{oman_spacelab_1987}; trains are studied by \cite{bertolini_determinants_2017,braccesi_development_2013}; road vehicles are widely studied as in \cite{wada_effect_2010,sugiura_analysing_2019,bronstein_visual_2020}.
Simulators are also widely studied because in such contexts the visual-vestibular conflict can be quite extreme due to vection; therefore several scientists as \cite{kennedy_simulator_1993,ohyama_autonomic_2007,zuzewicz_heart_2011} studied such topics.

\subsection[Autonomous driving]{Autonomous driving: challenge or opportunity?}

Autonomous driving is one of the most interesting growing technologies in the transportation field: it should bring safer roads avoiding human errors (\cite{maurer_safety_2016}) and should decrease the environmental impact of road transportation (\cite{fagnant_travel_2014}); the expected increase in passengers' productivity is instead debated within the scientific community (\cite{singleton_discussing_2019}): one of the greatest issues in being productive in an automated vehicle is indeed \gls{ms} (\cite{meyer_motion_2016}).

The proposed countermeasures for \gls{ms} in vehicles are mostly qualitative: \cite{diels_self-driving_2016} suggest to maximise the visual field of the occupants and to avoid rearward orientation of seats; \cite{wada_effect_2010} suggest to lean inward the head when approaching a turn to keep the head more aligned with the gravitoinertial acceleration; indication about future vehicle motion can be helpful as per \cite{kuiper_knowing_2020}, however, implementing head-up displays showing the planned trajectory as per \cite{feenstra_visual_2011}, might be infeasible for multi-seated cars.
A more quantitative approach has been proposed by \cite{sugiura_analysing_2019} using tilting seats to be more consistent compared to the head tilting strategy; however, similar approaches in tilting trains gave unclear results as per \cite{cohen_motion_2011,bertolini_determinants_2017}.

The idea of the authors is to try to exploit the peculiarities of autonomous vehicles: given a path, is it possible to define an optimal speed profile to minimise \gls{msi} and travel time for that mission?

To evaluate such capabilities, a numerical model of \gls{ms} is needed.

\subsection[Motion sickness models]{Numerical models for motion sickness: ISO approach and subjective vertical conflict theory}

Numerical models for \gls{ms} can be classified under two main categories: ISO-like approach and Bos\&Bles-like ones.
In \cite{lawther_prediction_1987} a method to estimate \gls{ms} by filtering the acceleration is presented; the integral of the squared filtered acceleration should be proportional to the \gls{msi}; \citeauthor[2631--1:1997]{iso_mechanical_1997} standard is based on Lawther and Griffin model.
In \cite{bos_modelling_1998}, a method is proposed modelling the mechanism of motion perception for vertical motion; three dimensional extensions of such modelling are proposed by \cite{braccesi_motion_2011,kamiji_modeling_2007}.
In \cite{braccesi_motion_2011} a comparison between their \emph{UniPG model} and the ISO one is showing that the UniPG model fits more consistently the literature data; furthermore, the Bos\&Bles derived models are capable of modelling the decrease in \gls{msi} when the stimulus stops, while the ISO model defines the \gls{msi} as a monotone positive value over time.

Despite these models are needed to assess the performance of the \gls{ms} reduction strategy implemented in this paper, are they really necessary to get a reduction in \gls{ms}?

\subsection[Optimisation aims]{Optimal approach to motion sickness reduction: a generalised quantitative approach}

Now that the issue of \gls{ms} has been introduced as well as some proposed countermeasures and numerical models, the goal of this research can be clearly defined: finding an optimal way to travel on a path trading off minimum travel time and minimal \gls{ms}.
Such technique, if proven to be effective, may be used to plan speed profile of an autonomous vehicle, to suggest an optimal speed profile to a human driver or to plan a more comfortable speed profile for bus line services or train services.
Such a solution can be combined with the techniques available in the literature to further reduce \gls{ms}; its main benefits compared to the one already available in the literature are:

\begin{itemize}
    \item it does not rely on passenger behaviour,
    \item it can be applied to already designed vehicles,
    \item it does not require complex active systems on the vehicle
    \item it does not require complex \gls{hmi} in the vehicles
\end{itemize}

Using the numerical model to assess the performance of the proposed algorithm allows to quantitatively evaluate the results only on an average estimate of the population.
In an actual implementation of this technique, such a limitation may be overcome by allowing the users to tune how strong the \gls{ms} reduction part of the optimisation should be; using a multi-objective cost function like the ones presented in this article lead to a straightforward implementation of this tuning by varying the \gls{ms} related cost in the cost function.

\section{Methods}

A simulation approach is used to compare different strategies to optimise speed profile on a given path. The results are analysed to understand if the modelling of \gls{ms} dynamics is the best approach to the trade-off between minimum travel time and \gls{ms} reduction or some strategy may be effective without explicit modelling of \gls{ms}.

The problem is implemented as a series of \gls{nmpc} problems representing the path through a series of sequential optimisations, leading to a formulation independent of path length.
An ideal representation of the simulated model has been assumed in the \gls{nmpc} problem, so the initial conditions of optimisation are imposed equal to states and input of the second step of the previous optimisation.
The NMPC has been implemented in MATLAB and solved using the \emph{fmincon} solver.
A shown in \cite{ohanlon_motion_1973}, \gls{ms} frequencies range from \SIrange{0.01}{1}{\hertz}, therefore the simulation step size has to be set to at least \SI{2}{\hertz} to get a proper description of \gls{ms} dynamics. To achieve such frequencies each optimisation space step size is defined as space travelled in \SI{0.5}{\second} at the optimisation initial velocity.
This approach leads to a different step size between low-velocity sharp turns and high-speed straights preventing an inefficient low step size due to the space transformation.

A test path is represented using a two-dimensional spline: the road used in this paper is a \SI{120}{\kilo\metre} part of an Italian motorway. The spline is created from an internet map service, transforming the latitude-longitudinal waypoints in North-East-Down coordinates leading to a flat x-y description of the test path. Several map services were compared, but different waypoints density between map services does not lead to significant changes in the resulting spline.
Neglecting the vertical contribution of the road, no vertical motion is present simplifying the formulation to a two-dimensional problem.
The path can be divided in two main sections: a first winding section characterised by a mountain crossing and a final straighter section; the first section is long nearly twice the later one.
Since automated driving will first used in motorways, the authors chose to use this scenario for their simulations; however the proposed approach is not limited to this scenario.

To get the acceleration of a vehicle moving on the path a vehicle model is needed; since the focus is not about a specific type of vehicle, but the optimisation of the given path, the simplest vehicle model is adopted: a point-mass model, which is fast to compute while being general for different vehicles.
Motion sickness is modelled using the \cite{braccesi_motion_2011} model; the UniPG model is computed for every strategy tested to get the evolution of the \gls{ms} incidence along the path, even if not used in the cost function.
To be more representative of a touring driving, the vehicle acceleration has been constrained to be less than \SI{0.3}{\g}. The jerk is also constrained to generate a smooth acceleration profile within \SI{\pm 3}{\metre\per\second\cubed}.
The vehicle is constrained on the spline, without any displacement in the normal direction, therefore the input of the model is only the longitudinal jerk; the lateral acceleration is computed from the longitudinal velocity and the path curvature.
To avoid issues due to the constant changing curvature during optimisation, the problem has been space transformed like in \cite{gao_spatial_2012}, so it is space integrated instead of time-integrated. This allows for a fixed curvature during the optimisation, even if it leads to a non-linear vehicle model; since the \gls{ms} model is non-linear in itself, non-linearity in the vehicle model was considered of minor concern and it was preferred to a changing curvature.
A detailed description of the model equations is given in \cite{certosini_preliminary_2019}.

The analysed strategies trading-off between \gls{ms} and travel time are implemented in several cost functions of the NPMC problem, their results are compared to understand which is the most effective in preventing \gls{ms} while keeping minimal travel time.
The cost functions analysed in this paper are described in table \ref{tab:CostFuncsDescr}, the equations are shown in table \ref{tab:CostFuncsEqs}, with the notation explained in table \ref{tab:Variables}.

\begin{table}
    \centering
    \caption{Description analysed cost functions}
    \label{tab:CostFuncsDescr}
    \settowidth\tymin{Adaptive MS cost}
    \begin{tabulary}{\linewidth}{CC} 
    \hline
    \textbf{Name} & \textbf{Description} \\ \hline
    Minimum time & Reference model that minimise travel time. Small cost on the input avoids excessive variation in the input. \\
    Jerk cost & Like minimum time, but with a high cost on the longitudinal jerk to reduce \gls{ms} \\
    Acceleration cost & Minimum time plus a cost on the acceleration modulus to reduce \gls{ms} \\
    MS cost & Minimum time plus an \gls{ms} cost using the instantaneous disturb of the UniPG model \\
    Adaptive MS cost & Similar to the previous one; the \gls{ms} cost is multiplied with the $\mathit{MSI}$ \\
    \hline
    \end{tabulary}
\end{table}

\begin{table}
    \centering
    \caption{Cost functions equations}
    \label{tab:CostFuncsEqs}
    \begin{tabulary}{\linewidth}{CC} 
    \hline
    \textbf{Name} & \textbf{Cost function} \\ \hline
    Minimum time & $C_{t} \dfrac{1}{v_{t}} + C_{u} j_{t}$ \\
    Jerk cost & $C_{t} \dfrac{1}{v_{t}} + C_{j} j_{t}$ \\
    Acceleration cost & $C_{t} \dfrac{1}{v_{t}} + C_{u} j_{t} + C_{a}\left( a_{t}^{2} + v_{t}^{2} * \rho\left(s\right) \right)$ \\
    MS cost & $C_{t} \dfrac{1}{v_{t}} + C_{u} j_{t} + C_{\mathit{MS}}h$ \\
    Adaptive MS cost & $C_{t} \dfrac{1}{v_{t}} + C_{u} j_{t} + C_{\mathit{MS}} h \mathit{MSI}$ \\
    \hline
    \end{tabulary}
\end{table}

\begin{table}
    \centering
    \caption{Cost functions variables}
    \label{tab:Variables}
    \settowidth\tymin{\textbf{Symbol}}
    \begin{tabulary}{\linewidth}{CC} 
    \hline
    \textbf{Symbol} & \textbf{Description} \\ \hline
    $C_{t}$ & Minimum time cost \\
    $C_{u}$ & Small input (jerk) cost to reduce input constraints violation\\
    $C_{j}$ & Big input (jerk) cost to reduce \gls{ms} \\
    $C_{a}$ & Acceleration cost \\
    $C_{\mathit{MS}}$ & Motion sickness cost \\
    $v_{t}$ & Tangential velocity \\
    $j_{t}$ & Tangential jerk (input) \\
    $a_{t}$ & Tangential acceleration \\
    $\rho$ & Path curvature \\
    $s$ & Space (independent variable) \\
    $h$ & \gls{ms} instantaneous disturb (as in \cite{braccesi_motion_2011}) \\
    $\mathit{MSI}$ & \gls{msi} (as in \cite{ohanlon_motion_1973} and computed as in \cite{bos_modelling_1998}) \\
    \hline
    \end{tabulary}
\end{table}

The analysed cost functions can be divided into model-free and model-based: the model-free approaches use some kinematic quantities like jerk and acceleration to reduce \gls{ms}, while the model-based ones use the UniPG model to reduce its incidence.
The first model-free cost function uses the jerk because it is a quantity related to comfort, so it is interesting to understand if it may be useful for this purpose.
The other one uses the acceleration: since the \gls{ms} models use often the acceleration as their input, as in \cite{ohanlon_motion_1973,lawther_prediction_1987,bos_modelling_1998,braccesi_motion_2011}, it is interesting to understand if reducing the input without modelling its dynamics may be sufficient to reduce the incidence of the discomfort.
A simple reduction of the maximum acceleration allowed without modifying the cost function has been proven in \cite{certosini_preliminary_2019} to be ineffective.
The model-based cost functions use the UniPG model to compute the instantaneous disturb variable $h$ and the current motion sickness incidence variable $\mathit{MSI}$ due to the evolution of the disturb: those are described in \cite{braccesi_motion_2011} respectively as a sort of non-linear-weighted filtering of the accelerations and as a sort of second-order leaking-integrator; for a detailed description of their mathematical formulation as block diagrams check \cite{bos_modelling_1998,braccesi_motion_2011}, while their formulation in terms of state equations is shown in \cite{certosini_preliminary_2019}.
The simple one has a cost $C_{\mathit{MS}}$ on the instantaneous disturb $h$: since $h$ is the most direct source of $\mathit{MSI}$, applying a cost on $h$ may be the most straightforward approach to reduce \gls{ms}; the drawback of such approach is that the weight in the cost function does not take in account the \gls{msi}, so even in a path too short to have a significant incidence the vehicle is slower than using the baseline cost function without significant advantage in terms of comfort.
A more advanced approach is to use both the instantaneous disturb and the incidence: the instantaneous disturb is weighted both by its cost $C_{\mathit{MS}}$ and $\mathit{MSI}$.
This approach, compared to the previous one, is capable of adapting itself to the road characteristics: if the road is too short to be relevant for \gls{ms}, a low $\mathit{MSI}$ will allow to a speed profile close to the minimum time one, when the path is long it will slow-down the vehicle during winding sections; due to this characteristic, the more advanced model has been named \emph{adaptive MS model}.

The performance of each cost function is assessed using the UniPG model comparing the maximum \gls{msi} obtained along the path against the travel time; the lower the time with the same maximum \gls{msi}, the more efficient the cost function is.
The results are also evaluated using the \citeauthor[2631--1:1997]{iso_mechanical_1997} model: coherent results would confirm their robustness to the metric used; to do that the fifth-order filter proposed in \cite{zuo_low_2003} is used to approximate the ISO filter.

\section{Results}

\begin{figure}[tbp]
    \centering
    \includegraphics[width = \columnwidth]{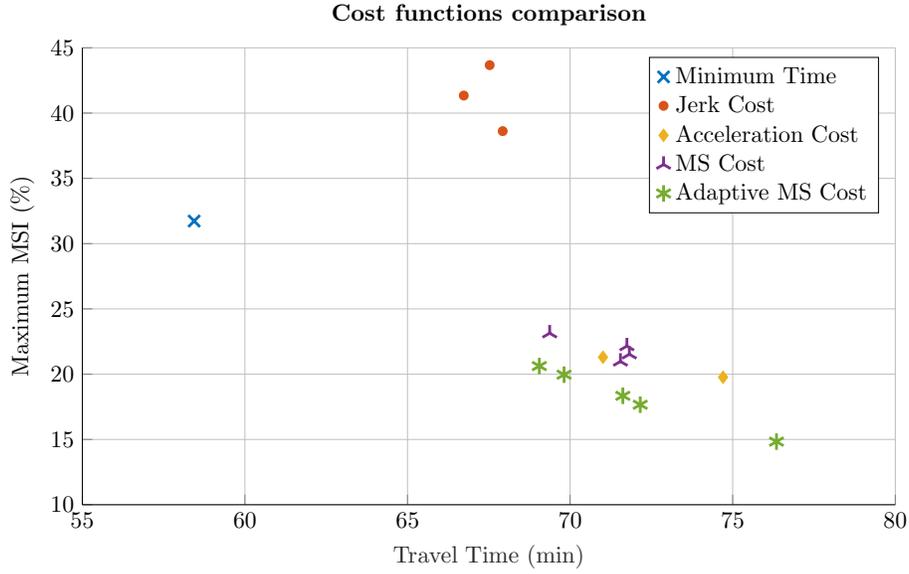}
    \caption{Maximum \gls{msi} over total travel time for the analysed cost functions}\label{fig:Pareto}
\end{figure}

The results are summarised in figure \ref{fig:Pareto}: the shorter the travel time and the lower the \gls{msi} the better; for a list of cost values for the various runs see table \ref{tab:Results}.

\begin{table}[tbp]
    \caption{Simulation results}
    \label{tab:Results}
    \begin{tabular}{cccc}
    \hline
    \textbf{Cost function}  & \textbf{Cost}         & \textbf{MSI max (\%)} & \textbf{Travel Time (min)}                        \\ \hline
    Minimum time            &                       & $31.7$                &               \SI{58}{\minute} \SI{26}{\second}   \\
    Jerk cost               & $C_{j}=40$            & $43.7$                & \SI{1}{\hour} \SI{07}{\minute} \SI{31}{\second}   \\
    Jerk cost               & $C_{j}=200$           & $41.3$                & \SI{1}{\hour} \SI{06}{\minute} \SI{43}{\second}   \\
    Jerk cost               & $C_{j}=1500$          & $38.6$                & \SI{1}{\hour} \SI{07}{\minute} \SI{55}{\second}   \\
    Acceleration cost       & $C_{a}=10$            & $21.3$                & \SI{1}{\hour} \SI{11}{\minute} \SI{00}{\second}   \\
    Acceleration cost       & $C_{a}=15$            & $19.8$                & \SI{1}{\hour} \SI{14}{\minute} \SI{42}{\second}   \\
    MS cost                 & $C_{\mathit{MS}}=75$  & $23.1$                & \SI{1}{\hour} \SI{09}{\minute} \SI{22}{\second}   \\
    MS cost                 & $C_{\mathit{MS}}=140$ & $21.0$                & \SI{1}{\hour} \SI{11}{\minute} \SI{32}{\second}   \\
    MS cost                 & $C_{\mathit{MS}}=170$ & $21.5$                & \SI{1}{\hour} \SI{11}{\minute} \SI{49}{\second}   \\
    MS cost                 & $C_{\mathit{MS}}=200$ & $22.2$                & \SI{1}{\hour} \SI{11}{\minute} \SI{44}{\second}   \\
    Adaptive MS cost        & $C_{\mathit{MS}}=24$  & $20.6$                & \SI{1}{\hour} \SI{09}{\minute} \SI{03}{\second}   \\
    Adaptive MS cost        & $C_{\mathit{MS}}=25$  & $20.0$                & \SI{1}{\hour} \SI{09}{\minute} \SI{48}{\second}   \\
    Adaptive MS cost        & $C_{\mathit{MS}}=27$  & $18.3$                & \SI{1}{\hour} \SI{11}{\minute} \SI{37}{\second}   \\
    Adaptive MS cost        & $C_{\mathit{MS}}=30$  & $17.7$                & \SI{1}{\hour} \SI{12}{\minute} \SI{09}{\second}   \\
    Adaptive MS cost        & $C_{\mathit{MS}}=40$  & $14.8$                & \SI{1}{\hour} \SI{16}{\minute} \SI{20}{\second}   \\ \hline
    \end{tabular}
\end{table}

The base model is, obviously the fastest, with a travel time of \SI{\approx 58}{\minute}; the maximum \gls{msi} of $\approx 32\%$ indicates quite a severe risk of emesis.

Despite its usage for comfort assessment, using the jerk to reduce \gls{ms} is ineffective.
The jerk cost function has the highest \gls{msi} and it is also quite slow.

The acceleration cost function proves to be very effective: it provides a significant reduction of \gls{msi} over the minimum time strategy and varying $C_{a}$ affects the trade-off between minimum time and low \gls{msi}.

An interesting result is the simple MS cost function one: just optimising $h$ in the optimisation is effective indeed, but it is not more effective than just trying to reduce the acceleration.
For the same \gls{msi} peak value, the travel time is usually higher than the acceleration run, therefore such technique is less efficient than the acceleration strategy.
For values of $C_{\mathit{MS}}$ in the $140-170-200$ range, the model has almost the same \gls{msi} and the same travel time, possibly due to some numerical issues for this cost function.

The adaptive MS cost function is the best performer of all those tested.
This cost function provides the lowest \gls{msi} among all and is substantially the most efficient between the three effective ones in reducing the \gls{msi}: comparing the runs with $C_{a}=10$, $C_{\mathit{MS}}=140$ and $C_{\mathit{MS}}=24$, the adaptive cost functions is faster by over two minutes compared to the others.
When looking for lower \gls{msi} values, the difference in time between the adaptive cost function and the acceleration cost function increases dramatically: lowering the maximum \gls{ms} from $21\%$ to $20\%$ the difference in time rise from \SI{\approx 2}{\minute} to \SI{\approx 5}{\minute}.

When looking at a travel time of \SI{\approx 71}{\minute}, the adaptive cost function reduces the maximum \gls{msi} to $\approx 18\%$ instead of the $\approx 21\%$ of the acceleration and simple MS cost functions.

\begin{figure}[tbp]
    \centering
    \includegraphics[width = \textwidth]{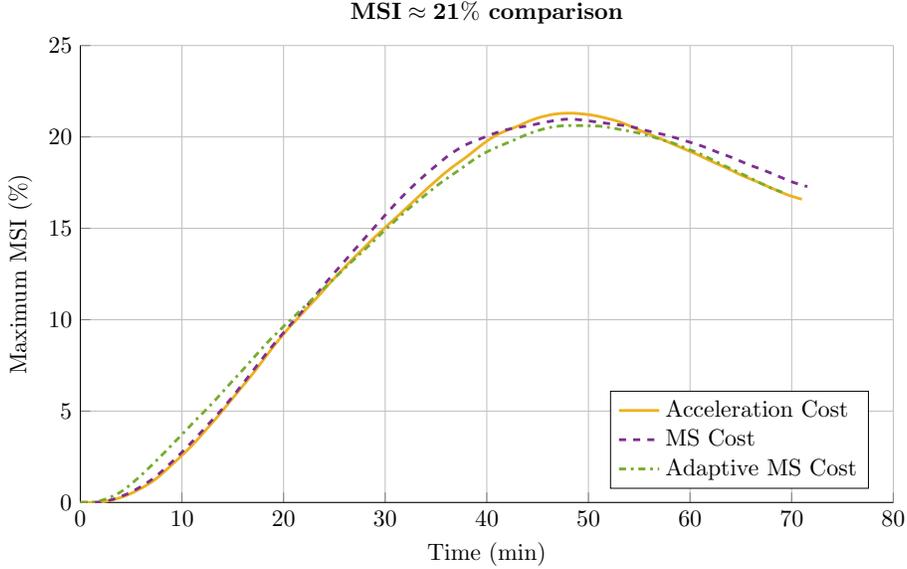}
    \caption{Comparison of three cost functions with similar maximum MSI}\label{fig:MSI21percent}
\end{figure}

\begin{figure}[tbp]
    \centering
    \includegraphics[width = \textwidth]{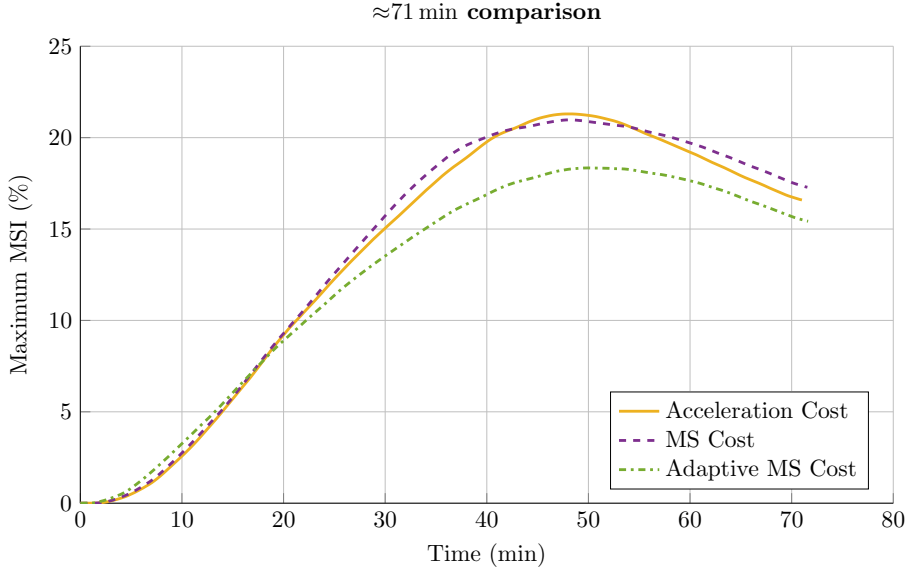}
    \caption{Comparison of three cost functions with similar travel time}\label{fig:MSI71min}
\end{figure}

Looking at the evolution of the \gls{msi} over time for runs with similar maximum \gls{msi} (figure \ref{fig:MSI21percent}), the difference between the adaptive cost function and the simpler MS one is interesting: at the beginning, the lower overall \gls{ms} cost ($C_{\mathit{MS}} \mathit{MSI}$) give higher \gls{msi} values, however, for higher \gls{msi}, the adaptation of the more complex cost function gives its benefits keeping the \gls{msi} below the simpler cost function.

When comparing runs of similar duration (figure \ref{fig:MSI71min}), it is clear how the adaptive cost function outperforms the others.

\section{Discussion}

The results show that the presented optimisation is effective in reducing passengers discomfort.
Introducing additional terms within the motion planning algorithm to increase passengers comfort is effective even without actively modelling the \gls{ms} dynamics as for the \emph{acceleration} cost function.
The best strategy for a simple implementation of an \gls{ms} reduction algorithm might be the \emph{acceleration cost}, but, for the best performance in terms of \gls{ms} reduction and minimal time, an \emph{adaptive MS} strategy is needed; the \emph{simple MS} strategy is not more effective than the \emph{acceleration cost} not justifying the additional burden of optimising the \gls{ms} dynamics.
The \gls{msi} data can be used as a feedback to the passengers: when using the \emph{acceleration cost}, the \gls{msi} needs to be computed as a separate task from the optimisation, while using an \emph{adaptive MS} strategy, the information is already available.

It is interesting to notice how the \emph{acceleration} cost function is effective while a further reduction of the maximum acceleration is not, as shown in \cite{certosini_preliminary_2019}.

\begin{figure}[tbp]
    \centering
    \includegraphics[width = \textwidth]{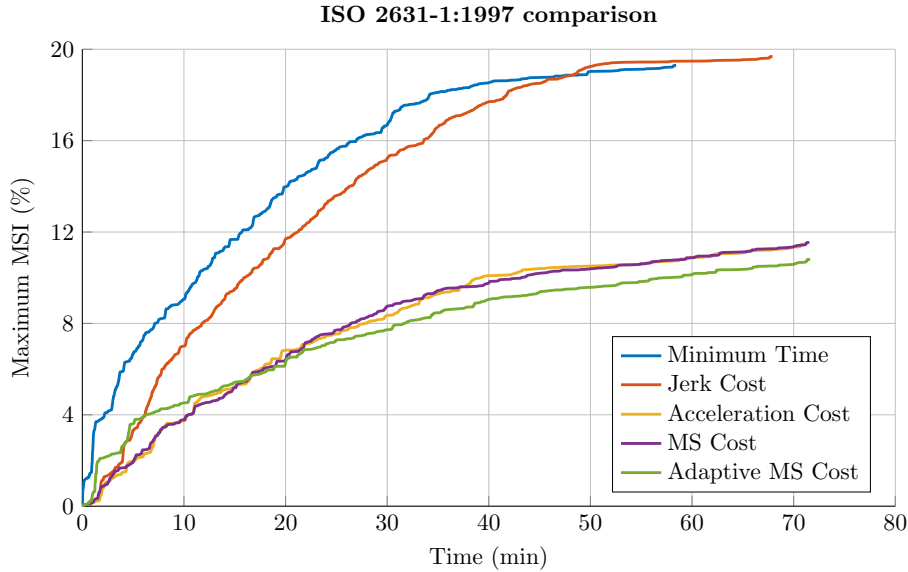}
    \caption{\gls{msi} estimated using the ISO-2631-1:1997 approach}\label{fig:ISOresults}
\end{figure}

Given the innate variability from person to person in the phenomenon of \gls{ms}, the numerical models are not so accurate; so it is interesting to compare the obtained results between the proposed cost functions using the ISO 2631:1997 approach to understand if the results are consistent and can be taken as a general improvement of comfort.
The results showed in figure \ref{fig:ISOresults} are consistent with the ones in figure \ref{fig:MSI71min}: the adaptive cost function is the most effective approach and the most efficient and the simple MS and acceleration cost functions have similar performance; it is interesting to notice the adaptation phase in the first minutes of the adaptive cost function.
Such coherency suggests that the proposed strategies do indeed offer an improvement in terms of \gls{msi} and that this improvement is not just a numerical trick due to the use of the same algorithm in optimisation and performance evaluation.

Despite leading to similar results, is shown in \cite{braccesi_motion_2011} that the UniPG model fits much better the experimental results of \cite{ohanlon_motion_1973,golding_frequency_1997} compared to \citeauthor[2631--1:1997]{iso_mechanical_1997}; according to this, results in figure \ref{fig:Pareto} are to be considered more reliable.

The usage of a Bos\&Bles-derived model for the computation of the \gls{msi} to be used in the cost function is key to model the decrease in \gls{msi} in straight stretches that follow winding sections, which is important in the adaptive cost function. Using ISO-like modelling would have led to an indefinitely increasing \gls{msi} over time, resulting in an unjustified slowing down of the vehicle on any straight stretch that follows tortuous sections, while it is a common experience that without or with small enough stimuli the sickness will eventually decrease.

The results of this research show that it is possible to improve passengers comfort and to reduce \gls{msi} by taking in account comfort-related terms in the optimisation during mission planning; it is shown how modelling their physiological dynamics is not mandatory to reduce \gls{msi}, but using a numerical model with the proposed adaptative cost function is essential to have good performance and efficiency.

The adaptive approach for mission speed profile optimisation is the main contribution of this paper, allowing for fine-tuning with small variations of the motion-sickness-related cost and the best performance in terms of both incidence reduction and minimal travel time.
The systematic comparison of the possible cost functions to reduce the \gls{msi} while minimising travel time is another contribution demonstrating that it is possible to obtain a reduction of the \gls{msi} without modelling its dynamics.

\section{Conclusion}

The presented study introduces a new approach to optimise speed profile on a given path: reducing \gls{msi} while still minimising travel time can significantly improve the acceptance of automated driving technologies.
The same approach can be used to improve comfort for human-driven cars advising the driver with the best speed or to improve services like intercity buses ones or even trains implementing \gls{ms} estimations while planning travel times; the approach proposed is not limited to autonomous vehicles given the simple vehicle model.

The proposed techniques can be integrated into motion planning algorithms for autonomous driving control to extend their benefits when the path definition is computed in real-time and not a priori defined like in this article.

To give the possibility to the users to interactively fine-tune the trade-off between \gls{ms} and travel time minimisation like in an adaptive cost function may greatly improve users' acceptance for autonomous vehicles and therefore improving their adoption, with the resulting benefits in terms of safety and environmental impact of transportation.

\printglossary[type=\acronymtype]

\bibliography{OptimSpeedMS}

\begin{thebibliography}{32}
\expandafter\ifx\csname natexlab\endcsname\relax\def\natexlab#1{#1}\fi
\providecommand{\url}[1]{\texttt{#1}}
\providecommand{\href}[2]{#2}
\providecommand{\path}[1]{#1}
\providecommand{\DOIprefix}{doi:}
\providecommand{\ArXivprefix}{arXiv:}
\providecommand{\URLprefix}{URL: }
\providecommand{\Pubmedprefix}{pmid:}
\providecommand{\doi}[1]{\href{http://dx.doi.org/#1}{\path{#1}}}
\providecommand{\Pubmed}[1]{\href{pmid:#1}{\path{#1}}}
\providecommand{\bibinfo}[2]{#2}
\ifx\xfnm\relax \def\xfnm[#1]{\unskip,\space#1}\fi
\bibitem[{Bertolini et~al.(2017)Bertolini, Durmaz, Ferrari, K{\"u}ffer, Lambert
  and Straumann}]{bertolini_determinants_2017}
\bibinfo{author}{Bertolini, G.}, \bibinfo{author}{Durmaz, M.A.},
  \bibinfo{author}{Ferrari, K.}, \bibinfo{author}{K{\"u}ffer, A.},
  \bibinfo{author}{Lambert, C.}, \bibinfo{author}{Straumann, D.},
  \bibinfo{year}{2017}.
\newblock \bibinfo{title}{Determinants of {{Motion Sickness}} in {{Tilting
  Trains}}: {{Coriolis}}/{{Cross}}-{{Coupling Stimuli}} and {{Tilt Delay}}}.
\newblock \bibinfo{journal}{Frontiers in Neurology} \bibinfo{volume}{8},
  \bibinfo{pages}{195}.
\newblock \DOIprefix\doi{10/ggqc6r}.
\bibitem[{Bos and Bles(1998)}]{bos_modelling_1998}
\bibinfo{author}{Bos, J.}, \bibinfo{author}{Bles, W.}, \bibinfo{year}{1998}.
\newblock \bibinfo{title}{Modelling motion sickness and subjective vertical
  mismatch detailed for vertical motions}.
\newblock \bibinfo{journal}{Brain Research Bulletin} \bibinfo{volume}{47},
  \bibinfo{pages}{537--542}.
\newblock \DOIprefix\doi{10/b9spsq}.
\bibitem[{Braccesi and Cianetti(2011)}]{braccesi_motion_2011}
\bibinfo{author}{Braccesi, C.}, \bibinfo{author}{Cianetti, F.},
  \bibinfo{year}{2011}.
\newblock \bibinfo{title}{Motion sickness. {{Part I}}: Development of a model
  for predicting motion sickness incidence}.
\newblock \bibinfo{journal}{International Journal of Human Factors Modelling
  and Simulation} \bibinfo{volume}{2}, \bibinfo{pages}{163}.
\newblock \DOIprefix\doi{10/fx2m69}.
\bibitem[{Braccesi et~al.(2013)Braccesi, Cianetti and
  Scaletta}]{braccesi_development_2013}
\bibinfo{author}{Braccesi, C.}, \bibinfo{author}{Cianetti, F.},
  \bibinfo{author}{Scaletta, R.}, \bibinfo{year}{2013}.
\newblock \bibinfo{title}{Development of a {{Methodology}} for the
  {{Evaluation}} of {{Motion Sickness Incidence}} in {{Railways}}},
  \bibinfo{publisher}{{ASME}}. p. \bibinfo{pages}{V013T14A045}.
\newblock \DOIprefix\doi{10/ggqc6s}.
\bibitem[{Bronstein et~al.(2020)Bronstein, Golding and
  Gresty}]{bronstein_visual_2020}
\bibinfo{author}{Bronstein, A.}, \bibinfo{author}{Golding, J.},
  \bibinfo{author}{Gresty, M.}, \bibinfo{year}{2020}.
\newblock \bibinfo{title}{Visual {{Vertigo}}, {{Motion Sickness}}, and
  {{Disorientation}} in {{Vehicles}}}.
\newblock \bibinfo{journal}{Seminars in Neurology} \bibinfo{volume}{40},
  \bibinfo{pages}{116--129}.
\newblock \DOIprefix\doi{10/ggqc6w}.
\bibitem[{Certosini et~al.(2019)Certosini, Papini, Capitani and
  Annicchiarico}]{certosini_preliminary_2019}
\bibinfo{author}{Certosini, C.}, \bibinfo{author}{Papini, L.},
  \bibinfo{author}{Capitani, R.}, \bibinfo{author}{Annicchiarico, C.},
  \bibinfo{year}{2019}.
\newblock \bibinfo{title}{Preliminary study for motion sickness reduction in
  autonomous vehicles: An {{MPC}} approach}.
\newblock \bibinfo{journal}{Procedia Structural Integrity}
  \bibinfo{volume}{24}, \bibinfo{pages}{127--136}.
\newblock \DOIprefix\doi{10/ggqc6v}.
\bibitem[{Cohen et~al.(2011)Cohen, Dai, Ogorodnikov, Laurens, Raphan,
  M{\"u}ller, Athanasios, Edmaier, Grossenbacher, Stadtm{\"u}ller, Brugger,
  Hauser and Straumann}]{cohen_motion_2011}
\bibinfo{author}{Cohen, B.}, \bibinfo{author}{Dai, M.},
  \bibinfo{author}{Ogorodnikov, D.}, \bibinfo{author}{Laurens, J.},
  \bibinfo{author}{Raphan, T.}, \bibinfo{author}{M{\"u}ller, P.},
  \bibinfo{author}{Athanasios, A.}, \bibinfo{author}{Edmaier, J.},
  \bibinfo{author}{Grossenbacher, T.}, \bibinfo{author}{Stadtm{\"u}ller, K.},
  \bibinfo{author}{Brugger, U.}, \bibinfo{author}{Hauser, G.},
  \bibinfo{author}{Straumann, D.}, \bibinfo{year}{2011}.
\newblock \bibinfo{title}{Motion sickness on tilting trains}.
\newblock \bibinfo{journal}{The FASEB Journal} \bibinfo{volume}{25},
  \bibinfo{pages}{3765--3774}.
\newblock \DOIprefix\doi{10/d38dcm}.
\bibitem[{Diels and Bos(2016)}]{diels_self-driving_2016}
\bibinfo{author}{Diels, C.}, \bibinfo{author}{Bos, J.E.}, \bibinfo{year}{2016}.
\newblock \bibinfo{title}{Self-driving carsickness}.
\newblock \bibinfo{journal}{Applied Ergonomics} \bibinfo{volume}{53},
  \bibinfo{pages}{374--382}.
\newblock \DOIprefix\doi{10/gf4sn5}.
\bibitem[{Diels et~al.(2016)Diels, Bos, Hottelart and
  Reilhac}]{meyer_motion_2016}
\bibinfo{author}{Diels, C.}, \bibinfo{author}{Bos, J.E.},
  \bibinfo{author}{Hottelart, K.}, \bibinfo{author}{Reilhac, P.},
  \bibinfo{year}{2016}.
\newblock \bibinfo{title}{Motion {{Sickness}} in {{Automated Vehicles}}: {{The
  Elephant}} in the {{Room}}}, in: \bibinfo{editor}{Meyer, G.},
  \bibinfo{editor}{Beiker, S.} (Eds.), \bibinfo{booktitle}{Road {{Vehicle
  Automation}} 3}. \bibinfo{publisher}{{Springer International Publishing}},
  \bibinfo{address}{{Cham}}, pp. \bibinfo{pages}{121--129}.
\newblock \DOIprefix\doi{10/dqsg}.
\bibitem[{Fagnant and Kockelman(2014)}]{fagnant_travel_2014}
\bibinfo{author}{Fagnant, D.J.}, \bibinfo{author}{Kockelman, K.M.},
  \bibinfo{year}{2014}.
\newblock \bibinfo{title}{The travel and environmental implications of shared
  autonomous vehicles, using agent-based model scenarios}.
\newblock \bibinfo{journal}{Transportation Research Part C: Emerging
  Technologies} \bibinfo{volume}{40}, \bibinfo{pages}{1--13}.
\newblock \DOIprefix\doi{10/f5wzwd}.
\bibitem[{Feenstra et~al.(2011)Feenstra, Bos and {van
  Gent}}]{feenstra_visual_2011}
\bibinfo{author}{Feenstra, P.}, \bibinfo{author}{Bos, J.},
  \bibinfo{author}{{van Gent}, R.}, \bibinfo{year}{2011}.
\newblock \bibinfo{title}{A visual display enhancing comfort by counteracting
  airsickness}.
\newblock \bibinfo{journal}{Displays} \bibinfo{volume}{32},
  \bibinfo{pages}{194--200}.
\newblock \DOIprefix\doi{10/fnp6k9}.
\bibitem[{Gao et~al.(2012)Gao, Gray, Frasch, Lin, Tseng, Hedrick and
  Borrelli}]{gao_spatial_2012}
\bibinfo{author}{Gao, Y.}, \bibinfo{author}{Gray, A.}, \bibinfo{author}{Frasch,
  J.V.}, \bibinfo{author}{Lin, T.}, \bibinfo{author}{Tseng, E.},
  \bibinfo{author}{Hedrick, J.K.}, \bibinfo{author}{Borrelli, F.},
  \bibinfo{year}{2012}.
\newblock \bibinfo{title}{Spatial {{Predictive Control}} for {{Agile
  Semi}}-{{Autonomous Ground Vehicles}}}, in: \bibinfo{booktitle}{Proceedings
  of the 11th {{International Symposium}} on {{Advanced Vehicle Control}}}.
\bibitem[{Golding et~al.(1997)Golding, Phil, Finch and
  Stott}]{golding_frequency_1997}
\bibinfo{author}{Golding, J.F.}, \bibinfo{author}{Phil, D.},
  \bibinfo{author}{Finch, M.I.}, \bibinfo{author}{Stott, J.R.R.},
  \bibinfo{year}{1997}.
\newblock \bibinfo{title}{Frequency effect of 0.35-1.{{0Hz}} horizontal
  translational oscillation on motion sickness and the somatogravic illusion}.
\newblock \bibinfo{journal}{Aviation, Space, and Environmental Medicine}
  \bibinfo{volume}{68}, \bibinfo{pages}{396--402}.
\bibitem[{Irwin(1881)}]{irwin_pathology_1881}
\bibinfo{author}{Irwin, J.}, \bibinfo{year}{1881}.
\newblock \bibinfo{title}{The pathology of sea-sickness}.
\newblock \bibinfo{journal}{The Lancet} \bibinfo{volume}{118},
  \bibinfo{pages}{907--909}.
\newblock \DOIprefix\doi{10/fnb6xz}.
\bibitem[{{ISO}(1997)}]{iso_mechanical_1997}
\bibinfo{author}{{ISO}}, \bibinfo{year}{1997}.
\newblock \bibinfo{title}{Mechanical Vibration and Shock \textemdash{}
  {{Evaluation}} of Human Exposure to Whole-Body Vibration \textemdash{}
  {{Part}} 1: {{General}} Requirements}.
\newblock \bibinfo{type}{{{ISO Standard}}} \bibinfo{number}{2631-1}.
  {International Organization for Standardization}.
  \bibinfo{address}{{Geneva}}.
\bibitem[{Kamiji et~al.(2007)Kamiji, Kurata, Wada and
  Doi}]{kamiji_modeling_2007}
\bibinfo{author}{Kamiji, N.}, \bibinfo{author}{Kurata, Y.},
  \bibinfo{author}{Wada, T.}, \bibinfo{author}{Doi, S.}, \bibinfo{year}{2007}.
\newblock \bibinfo{title}{Modeling and validation of carsickness mechanism},
  in: \bibinfo{booktitle}{{{SICE Annual Conference}} 2007},
  \bibinfo{publisher}{{IEEE}}, \bibinfo{address}{{Takamatsu, Japan}}. pp.
  \bibinfo{pages}{1138--1143}.
\newblock \DOIprefix\doi{10/fmvp69}.
\bibitem[{Kennedy et~al.(1993)Kennedy, Lane, Berbaum and
  Lilienthal}]{kennedy_simulator_1993}
\bibinfo{author}{Kennedy, R.S.}, \bibinfo{author}{Lane, N.E.},
  \bibinfo{author}{Berbaum, K.S.}, \bibinfo{author}{Lilienthal, M.G.},
  \bibinfo{year}{1993}.
\newblock \bibinfo{title}{Simulator {{Sickness Questionnaire}}: {{An Enhanced
  Method}} for {{Quantifying Simulator Sickness}}}.
\newblock \bibinfo{journal}{The International Journal of Aviation Psychology}
  \bibinfo{volume}{3}, \bibinfo{pages}{203--220}.
\newblock \DOIprefix\doi{10/bbh54v}.
\bibitem[{Kuiper et~al.(2020)Kuiper, Bos, Diels and
  Schmidt}]{kuiper_knowing_2020}
\bibinfo{author}{Kuiper, O.X.}, \bibinfo{author}{Bos, J.E.},
  \bibinfo{author}{Diels, C.}, \bibinfo{author}{Schmidt, E.A.},
  \bibinfo{year}{2020}.
\newblock \bibinfo{title}{Knowing what's coming: {{Anticipatory}} audio cues
  can mitigate motion sickness}.
\newblock \bibinfo{journal}{Applied Ergonomics} \bibinfo{volume}{85},
  \bibinfo{pages}{103068}.
\newblock \DOIprefix\doi{10/ggp88p}.
\bibitem[{Lawther and Griffin(1986)}]{lawther_motion_1986}
\bibinfo{author}{Lawther, A.}, \bibinfo{author}{Griffin, M.J.},
  \bibinfo{year}{1986}.
\newblock \bibinfo{title}{The motion of a ship at sea and the consequent motion
  sickness amongst passengers}.
\newblock \bibinfo{journal}{Ergonomics} \bibinfo{volume}{29},
  \bibinfo{pages}{535--552}.
\newblock \DOIprefix\doi{10/fk9tsh}.
\bibitem[{Lawther and Griffin(1987)}]{lawther_prediction_1987}
\bibinfo{author}{Lawther, A.}, \bibinfo{author}{Griffin, M.J.},
  \bibinfo{year}{1987}.
\newblock \bibinfo{title}{Prediction of the incidence of motion sickness from
  the magnitude, frequency, and duration of vertical oscillation}.
\newblock \bibinfo{journal}{The Journal of the Acoustical Society of America}
  \bibinfo{volume}{82}, \bibinfo{pages}{957--966}.
\newblock \DOIprefix\doi{10/cw29cs}.
\bibitem[{O'Hanlon and McCauley(1973)}]{ohanlon_motion_1973}
\bibinfo{author}{O'Hanlon, J.F.}, \bibinfo{author}{McCauley, M.E.},
  \bibinfo{year}{1973}.
\newblock \bibinfo{title}{Motion {{Sickness Incidence}} as a {{Function}} of
  the {{Frequency}} and {{Acceleration}} of {{Vertical Sinusoidal Motion}}}.
\newblock \bibinfo{type}{Technical Report} \bibinfo{number}{AD0768215}. {Office
  of Naval Research}.
\bibitem[{Ohyama et~al.(2007)Ohyama, Nishiike, Watanabe, Matsuoka, Akizuki,
  Takeda and Harada}]{ohyama_autonomic_2007}
\bibinfo{author}{Ohyama, S.}, \bibinfo{author}{Nishiike, S.},
  \bibinfo{author}{Watanabe, H.}, \bibinfo{author}{Matsuoka, K.},
  \bibinfo{author}{Akizuki, H.}, \bibinfo{author}{Takeda, N.},
  \bibinfo{author}{Harada, T.}, \bibinfo{year}{2007}.
\newblock \bibinfo{title}{Autonomic responses during motion sickness induced by
  virtual reality}.
\newblock \bibinfo{journal}{Auris Nasus Larynx} \bibinfo{volume}{34},
  \bibinfo{pages}{303--306}.
\newblock \DOIprefix\doi{10/bbgrjr}.
\bibitem[{Oman(1987)}]{oman_spacelab_1987}
\bibinfo{author}{Oman, C.M.}, \bibinfo{year}{1987}.
\newblock \bibinfo{title}{Spacelab experiments on space motion sickness}.
\newblock \bibinfo{journal}{Acta Astronautica} \bibinfo{volume}{15},
  \bibinfo{pages}{55--66}.
\newblock \DOIprefix\doi{10/chs3mb}.
\bibitem[{Reason and Brand(1975)}]{reason_motion_1975}
\bibinfo{author}{Reason, J.T.}, \bibinfo{author}{Brand, J.J.},
  \bibinfo{year}{1975}.
\newblock \bibinfo{title}{Motion Sickness}.
\newblock \bibinfo{publisher}{{Academic Press}}, \bibinfo{address}{{London, New
  York}}.
\bibitem[{Singleton(2019)}]{singleton_discussing_2019}
\bibinfo{author}{Singleton, P.A.}, \bibinfo{year}{2019}.
\newblock \bibinfo{title}{Discussing the ``positive utilities'' of autonomous
  vehicles: Will travellers really use their time productively?}
\newblock \bibinfo{journal}{Transport Reviews} \bibinfo{volume}{39},
  \bibinfo{pages}{50--65}.
\newblock \DOIprefix\doi{10/ggp9z2}.
\bibitem[{Sugiura et~al.(2019)Sugiura, Wada, Nagata, Sakai and
  Sato}]{sugiura_analysing_2019}
\bibinfo{author}{Sugiura, T.}, \bibinfo{author}{Wada, T.},
  \bibinfo{author}{Nagata, T.}, \bibinfo{author}{Sakai, K.},
  \bibinfo{author}{Sato, Y.}, \bibinfo{year}{2019}.
\newblock \bibinfo{title}{Analysing {{Effect}} of {{Vehicle Lean Using
  Cybernetic Model}} of {{Motion Sickness}}}.
\newblock \bibinfo{journal}{IFAC-PapersOnLine} \bibinfo{volume}{52},
  \bibinfo{pages}{311--316}.
\newblock \DOIprefix\doi{10/ggp88q}.
\bibitem[{Turner et~al.(2000)Turner, Griffin and
  Holland}]{turner_airsickness_2000}
\bibinfo{author}{Turner, M.}, \bibinfo{author}{Griffin, M.J.},
  \bibinfo{author}{Holland, I.}, \bibinfo{year}{2000}.
\newblock \bibinfo{title}{Airsickness and aircraft motion during short-haul
  flights}.
\newblock \bibinfo{journal}{Aviation, Space, and Environmental Medicine}
  \bibinfo{volume}{71}, \bibinfo{pages}{1181--1189}.
\bibitem[{Wada et~al.(2010)Wada, Fujisawa, Imaizumi, Kamiji and
  Doi}]{wada_effect_2010}
\bibinfo{author}{Wada, T.}, \bibinfo{author}{Fujisawa, S.},
  \bibinfo{author}{Imaizumi, K.}, \bibinfo{author}{Kamiji, N.},
  \bibinfo{author}{Doi, S.}, \bibinfo{year}{2010}.
\newblock \bibinfo{title}{Effect of {{Driver}}'s {{Head Tilt Strategy}} on
  {{Motion Sickness Incidence}}}.
\newblock \bibinfo{journal}{IFAC Proceedings Volumes} \bibinfo{volume}{43},
  \bibinfo{pages}{192--197}.
\newblock \DOIprefix\doi{10/dkfpqf}.
\bibitem[{Winkle(2016)}]{maurer_safety_2016}
\bibinfo{author}{Winkle, T.}, \bibinfo{year}{2016}.
\newblock \bibinfo{title}{Safety {{Benefits}} of {{Automated Vehicles}}:
  {{Extended Findings}} from {{Accident Research}} for {{Development}},
  {{Validation}} and {{Testing}}}, in: \bibinfo{editor}{Maurer, M.},
  \bibinfo{editor}{Gerdes, J.C.}, \bibinfo{editor}{Lenz, B.},
  \bibinfo{editor}{Winner, H.} (Eds.), \bibinfo{booktitle}{Autonomous
  {{Driving}}}. \bibinfo{publisher}{{Springer Berlin Heidelberg}},
  \bibinfo{address}{{Berlin, Heidelberg}}, pp. \bibinfo{pages}{335--364}.
\newblock \DOIprefix\doi{10/dqsh}.
\bibitem[{Young(1978)}]{young_visually_1978}
\bibinfo{author}{Young, L.R.}, \bibinfo{year}{1978}.
\newblock \bibinfo{title}{Visually induced motion in flight simulation}, in:
  \bibinfo{booktitle}{{{AGARD Symposioum}} on Flight Simulation},
  \bibinfo{address}{{Brussels}}.
\bibitem[{Zuo and Nayfeh(2003)}]{zuo_low_2003}
\bibinfo{author}{Zuo, L.}, \bibinfo{author}{Nayfeh, S.}, \bibinfo{year}{2003}.
\newblock \bibinfo{title}{Low order continuous-time filters for approximation
  of the {{ISO}} 2631-1 human vibration sensitivity weightings}.
\newblock \bibinfo{journal}{Journal of Sound and Vibration}
  \bibinfo{volume}{265}, \bibinfo{pages}{459--465}.
\newblock \DOIprefix\doi{10/dgpmtz}.
\bibitem[{Zu{\.z}ewicz et~al.(2011)Zu{\.z}ewicz, Saulewicz, Konarska and
  Kaczorowski}]{zuzewicz_heart_2011}
\bibinfo{author}{Zu{\.z}ewicz, K.}, \bibinfo{author}{Saulewicz, A.},
  \bibinfo{author}{Konarska, M.}, \bibinfo{author}{Kaczorowski, Z.},
  \bibinfo{year}{2011}.
\newblock \bibinfo{title}{Heart rate variability and motion sickness during
  forklift simulator driving}.
\newblock \bibinfo{journal}{International journal of occupational safety and
  ergonomics: JOSE} \bibinfo{volume}{17}, \bibinfo{pages}{403--410}.
\newblock \DOIprefix\doi{10/ggqc6t}.

\end{thebibliography}
\bibliographystyle{elsarticle-harv}

\end{document}